\documentclass[a4paper,fleqn,usenatbib]{mnras}

\usepackage{newtxtext,newtxmath}

\usepackage[T1]{fontenc}
\usepackage{ae,aecompl}

\usepackage{amsmath}	
\usepackage{amssymb}	
\usepackage{bm}
\usepackage{here}
\usepackage[pdftex]{graphicx}

\usepackage{color}

\title[Iron line predictions for the wind in H~1743-322]
{The thermal-radiative wind in low mass X-ray binary H~1743-322: 
\newline 
II. iron line predictions from Monte Carlo radiation transfer
}

\author[Tomaru et al.]{
\thanks{E-mail: ryota.tomaru@ipmu.jp}
Ryota Tomaru,$^{1,2}$
Chris Done,$^{3,2}$
Ken Ohsuga,$^{4}$
Hirokazu Odaka$^{1,2}$ and
\newauthor
Tadayuki Takahashi$^{2,1}$ 
\\
$^{1}$Department of Physics, The University of Tokyo, 7-3-1 Hongo, Bunkyo, Tokyo 113-0033, Japan\\
$^{2}$Kavli Institute for the Physics and Mathematics of the Universe (WPI), University of Tokyo, Kashiwa 277-8583, Japan  \\
$^{3}$ Centre for Extragalactic Astronomy, Department of Physics, Durham University, South Road, Durham, DH1 4ED, UK\\
$^{4}$Center for Computational Sciences, University of Tsukuba, 1-1-1- Ten-nodai, Tsukuba, Ibaraki, 305-8577, Japan \\
}

\date{Accepted XXX. Received YYY; in original form ZZZ}

\pubyear{2015}

\begin{document}
\label{firstpage}
\pagerange{\pageref{firstpage}--\pageref{lastpage}}
\maketitle

\begin{abstract}

We show the best current simulations of the absorption and emission features predicted from thermal-radiative winds produced from X-ray illumination of the outer accretion disc in binary systems. We use the density and velocity structure derived from a radiation hydrodynamic code as input to a Monte-Carlo radiation transport calculation. The initial conditions are matched to those of the black hole binary system H1743-322 in its soft, disc dominated state, where wind features are seen in {\it Chandra} grating data. Our simulation fits well to the observed line profile, showing that these physical wind models can be the origin of the absorption features seen, rather than requiring a magnetically driven wind. 
We show how the velocity structure is the key observable discriminator between magnetic and thermal winds. Magnetic winds are faster at smaller radii, whereas thermal winds transition to a static atmosphere at smaller radii. New data from {\it XRISM} (due for launch Jan 2022) will give an unprecedented view of the physics of the wind launch and acceleration processes, but the existence of static atmospheres in small disc systems already rules out magnetic winds which assume self-similar magnetic fields from the entire disc as the origin of the absorption features seen. 
\end{abstract}

\section{introduction}

Accretion disc winds play an important role in the physics of accretion onto compact objects, carrying 
significant mass, angular momentum and/or kinetic energy away from the system. 
These winds are ubiquitous across the mass scale, from protostars (e.g.\citealt{Bjerkeli2016}) through to stellar mass compact objects (both black holes and neutron stars) in 
X-ray binaries (e.g. \citealt{Ponti2012, DiazTrigo2016}) and the supermassive black holes in active 
galactic nuclei (AGN) at the centre of galaxies (e.g. \citealt{Nardini2015}).
The stellar mass black hole binaries (BHB) especially offer a window into the poorly known physics of 
how the outflows are launched and accelerated as they vary dramatically in both spectral shape and 
brightness on easily observed timescales of few days to few months. 
This allows us to study how the wind responds to the changing accretion flow, constraining its nature. This is especially marked during the transition from thermal disc dominated spectra
(soft state) \citep{Shakura1973} to the Comptonized spectra (hard state), most likely from a
hot, geometrically thick, optically thin type accretion flow \citep{Ichimaru1977,Narayan1994} which is also the base of the jet seen in this state \citep{Fender2004}.

The existence of winds are shown by blueshifted absorption lines from highly ionised ions.
These are only seen in soft state but not in hard state \citep{Ponti2012}, anti-correlated with the radio jet which is seen in the hard state but not in the soft. This was thought to be evidence that the wind was magnetically driven by the same field as was responsible for the jet, but in 
a different geometric configuration \citep{Miller2012}. However, in \citet[hereafter Paper I]{Tomaru2019} we show instead that 
thermally driven winds can explain this switch (see also \citealt{Done2018, Shidatsu2019}).
Thermal driving produces a wind by irradiation from the central source heating the surface of accretion disc up to the Compton temperature ($T_\text{IC} \sim 10^7 -10^8 \text{K}$), which is hot enough for its thermal energy to overcome the gravity at large radii.
The characteristic radius at which the wind can be launched is called the Compton radius, defined by $R_\text{IC} = \mu m_p GM/kT_\text{IC}\sim 10^{5}-10^{6} R_\text{g}$ \citep{Begelman1983a}. 
Paper I show the first modern radiation hydrodynamic simulations of thermal (and thermal-radiative) winds designed to investigate the switch in wind properties between the hard and soft states changing illumination spectra.
These simulations were tailored to the BHB system H1743-332, where there is {\it Chandra} high resolution data in both states giving detailed spectral information on the wind or its absence \citep{Miller2012,Shidatsu2019}.
They incorporate radiation force on the electrons, both bound and free, as they show that this is important factor driving the escape of the thermal wind in the fairly high Eddington fraction ($L/L_\text{Edd}\sim 0.2-0.3$), fairly low Compton temperature ($T_\text{IC}\sim 0.1\times 10^8$~K) characteristics of the soft state. The only other modern hydrodynamic simulation of thermal winds 
(e.g. \citealt{Luketic2010, Higginbottom2015, Higginbottom2018} ) have not included radiation pressure, which is important in setting the velocity structure for $L\ge 0.3L_\text{Edd}$ as required here (Paper I). 

Paper I shows that the thermal-radiative wind calculation match well to the observed wind properties (column densities of the absorption lines of Fe {\scriptsize XXV} and Fe {\scriptsize XXVI}, together with the velocity) derived from {\it Chandra} grating data. Here we take the wind density and velocity field produced in the radiation hydrodynamic calculations and use this as input to a Monte-Carlo radiation transport code, {\sc monaco} \citep{Odaka2011} to make more detailed calculations and simulations of the iron line emission and absorption line profiles. We show that these match well to current data, and use these to predict the signatures of  
thermal-radiative winds in the upcomming high resolution observations using {\it XRISM}/Resolve (due for launch Jan 2022). Additionally, we show how this X-ray heated disc atmosphere/wind can strongly constrain the alternative magnetic wind models. 

\section{Method}

\begin{figure}
    \centering
    \includegraphics[width=0.8\hsize]{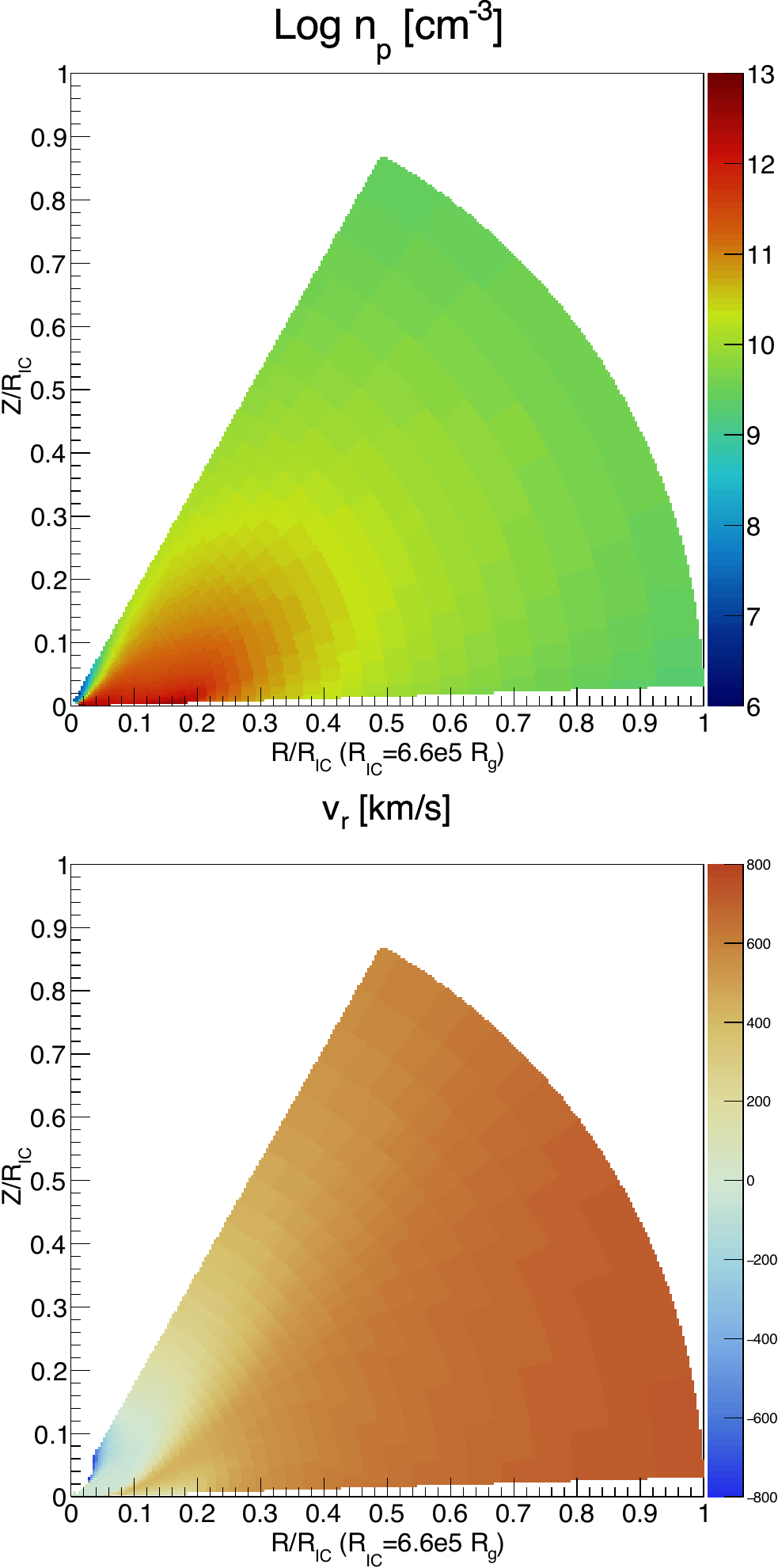}
    \caption{Distribution of the density (top), outflow velocity (bottom). The dark blue region of velocity map means negative velocity }
    \label{fig:hydro_out}
\end{figure}
\begin{figure*}
    \centering
    \includegraphics[width=0.9\hsize]{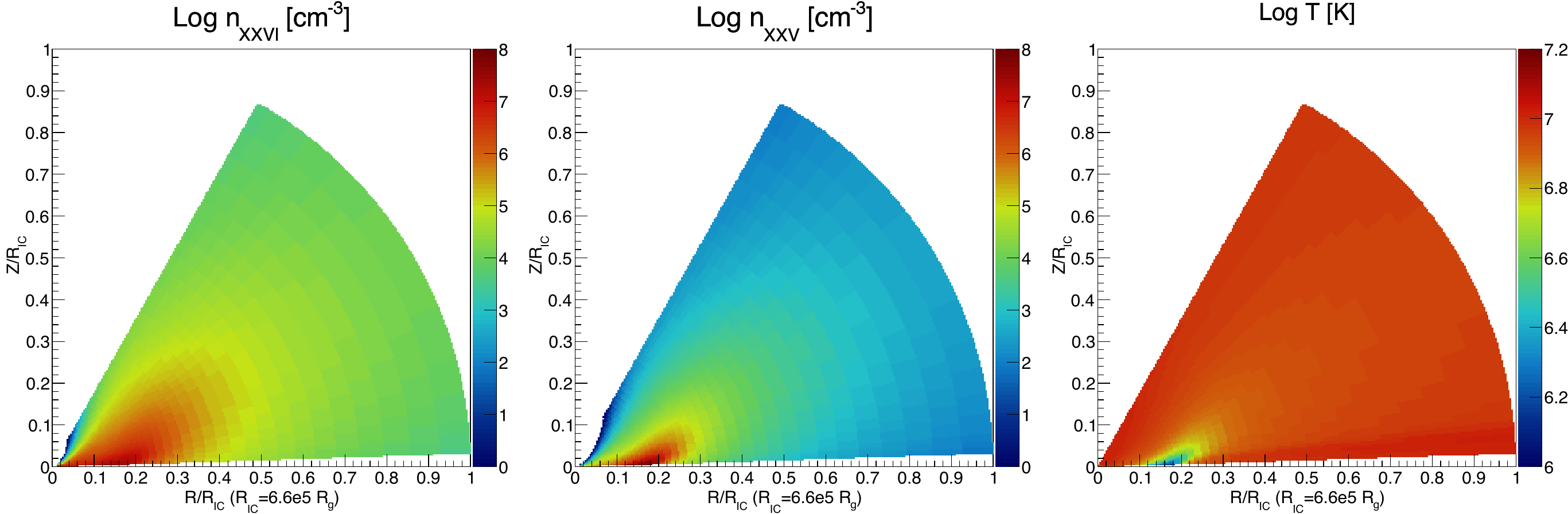}
    \caption{Distributions of ion density Fe {\scriptsize XXVI} (left), Fe {\scriptsize XXV} (middle) and temperature (right) calculated by {\sc cloudy}}
    \label{fig:cloudy_out}
\end{figure*}

\begin{figure*}
    \centering
    \includegraphics[width=0.9\hsize]{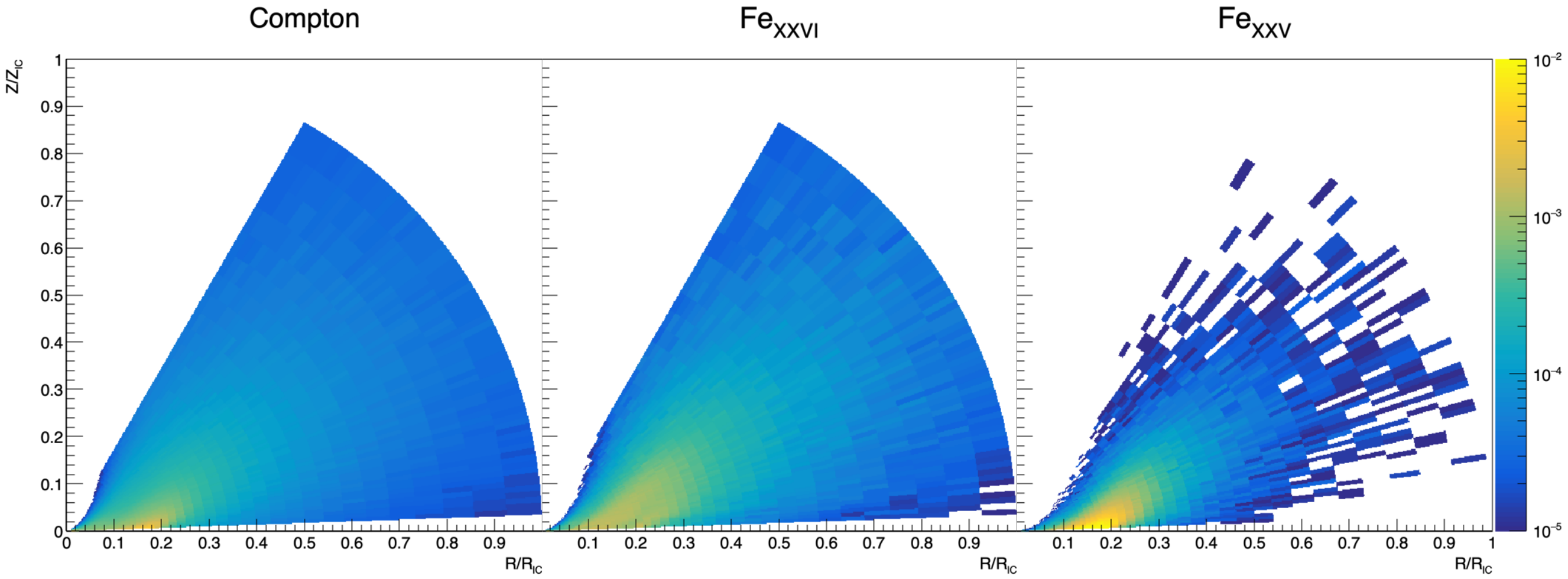}
    \caption{The position distribution of reprocessed emission from Compton scattering (left), Fe {\sc xxvi} (middle), and Fe {\sc xxv} (right).
    The colour shows that the ratio of the local counts to the total counts of these reprocessed emissions.
    }
    \label{fig:dist emission}
\end{figure*}
H1743-322 is one of the BHB which shows wind features during its outbursts \citep{Miller2006b}.
Its binary parameters are not well known, but in Paper I we show a simulation in section (4.2) which is most consistent with the overall wind properties seen in the {\it Chandra} grating data. 
This uses black hole mass $M = 7~M_\odot$, inclination $i = 75 \pm 3 ^{\circ}$ and distance $D = 8.5~\text{kpc}$ \citep{Steiner2012} with outer disc radius $R_\text{out}=1.2\times 10^5~R_\text{g}$.
The simulation includes realistic heating/cooling rates of the plasma calculated from the {\sc cloudy} \citep{Ferland2003} photo-ionisation code for the observed broadband spectrum.
It also incorporates the expected shadowing effect from the X-ray heated atmosphere above the inner disc. This shields the outer disc from direct irradiation until the flared shape of the outer disc rises above the shadow \citet{Begelman1983b}.

We use the density and velocity results of this radiation hydrodynamics simulation (Fig. \ref{fig:hydro_out}) as input to the radiative transfer Monte-Carlo code {\sc monaco} \citep{Odaka2011}. We rebin these via bi-linear interpolation 
to reduce the simulation grid from $(N_r, N_\theta) = (120, 240)$ to $ (60, 120)$ and only include the region of $\theta = 30-88^{\circ}$ i.e. removing the low density polar region and the high density disc region.
We explicitly include azimuth, so the 
total grids we use are 60 (radial) and  51 (polar) and 32 (azimuth).

{\sc monaco} uses the Geant4 
toolkit library \citep{Agostinelli2003} for photon tracking in an arbitrary three-dimensional geometry, but has its own 
modules handling photon interactions \citet{Watanabe2006,Odaka2011} so that it can 
treat the interactions 
such as photo-ionisation or photo-excitation, 
and photons generated via recombination and atomic de-excitation. 
Scattering by free electrons also is taken into account. 
The code also handles the Doppler shift of the absorption cross section from the velocity structure of the material. 
This cross section is calculated for the photon energy in the comoving frame and Lorentz tansformed back into the rest frame.
The Doppler broadening of temperature and turbulent motion is also considered.
The atomic data for all the transitions considered are tabulated in Mizumoto et al in prep, and  the overall iron abundance is $A_\text{Fe}=3.3\times 10^{-5}$.

{\sc monaco} also requires the distribution of ion populations and temperature in addition to density and velocity. We obtain these 
more accurately than is possible in the more approximate approach of the 
radiation hydrodynamic code by solving the one dimensional radiation transfer along line of sight using  {\sc cloudy}.
We chain {\sc cloudy} radially through the density structure and use the output spectrum of inner grid as the input spectrum to the next grid. 
The initial source spectrum is the same as that used to make the radiation hydrodynamic simulation i.e.  is the spectrum of  H1743-322 in its soft state at the time of the {\it Chandra} grating observation of \citet{Miller2006b}.

Fig. \ref{fig:cloudy_out} shows the resulting distribution of the density of Fe {\scriptsize XXVI}, Fe {\scriptsize XXV} (left, middle) and temperature (right). 
The density of both H and He-like Fe is highest near to the mid-plane 
because this is where the wind density is highest (see Fig.\ref{fig:hydro_out}. H-like iron is produced interior to He-like iron as it is formed at a higher ionisation parameter. The temperature in most of the wind region is the Compton temperature ($T_\text{IC} = 0.1\times 10^8 \text{K}$), except for the highest density region near to the disc surface. 

We generate photons isotropically from the centre 
over the energy range $6.5-8.5$~keV with 1~eV resolution (3000 bins). 
The total number of input photons is $1.4\times 10^8$. 
{\sc monaco} tracks all the interactions of these photons from their creation to eventual escape.

\section{Iron line Emission and Absorption line Profiles}

\begin{figure*}
    \centering
    \includegraphics[width=0.9 \hsize]{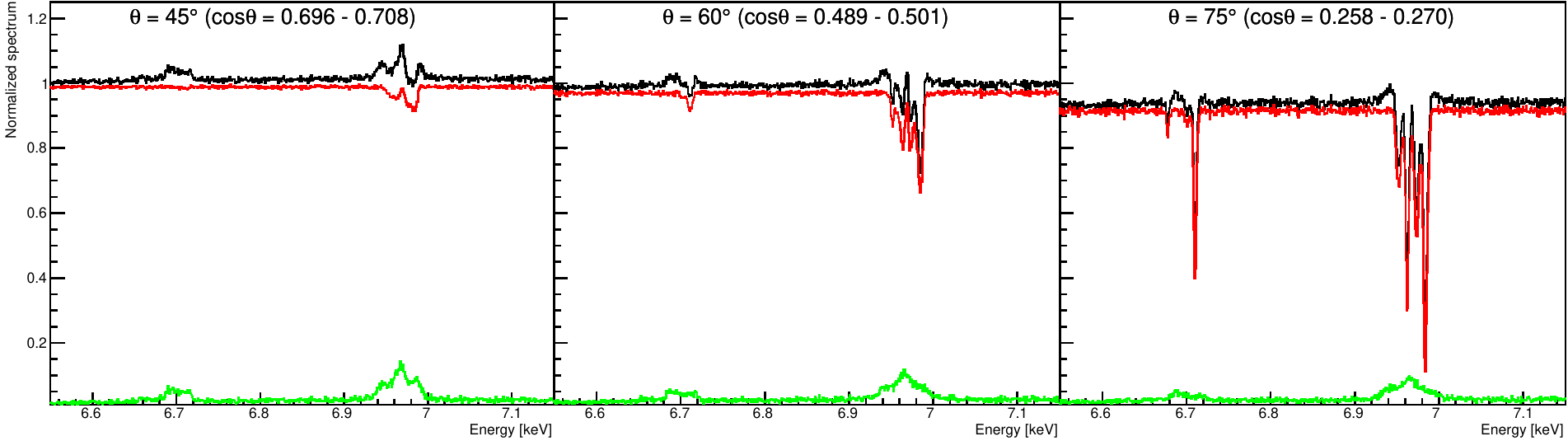}
    \caption{The results using the density/velocity structure from the hydrodynamic simulations (i.e. no additional turbulent velocity). Each panel shows the calculated spectra (black) seen at inclination of $45^\circ, 60^\circ$ and $75^\circ$, normalised by the incident spectrum. The transmitted (red) and scattered (green) components are also shown.}
    \label{fig:vt_zero_each}
\end{figure*}

\begin{figure*}
    \centering
    \includegraphics[width=0.9 \hsize]{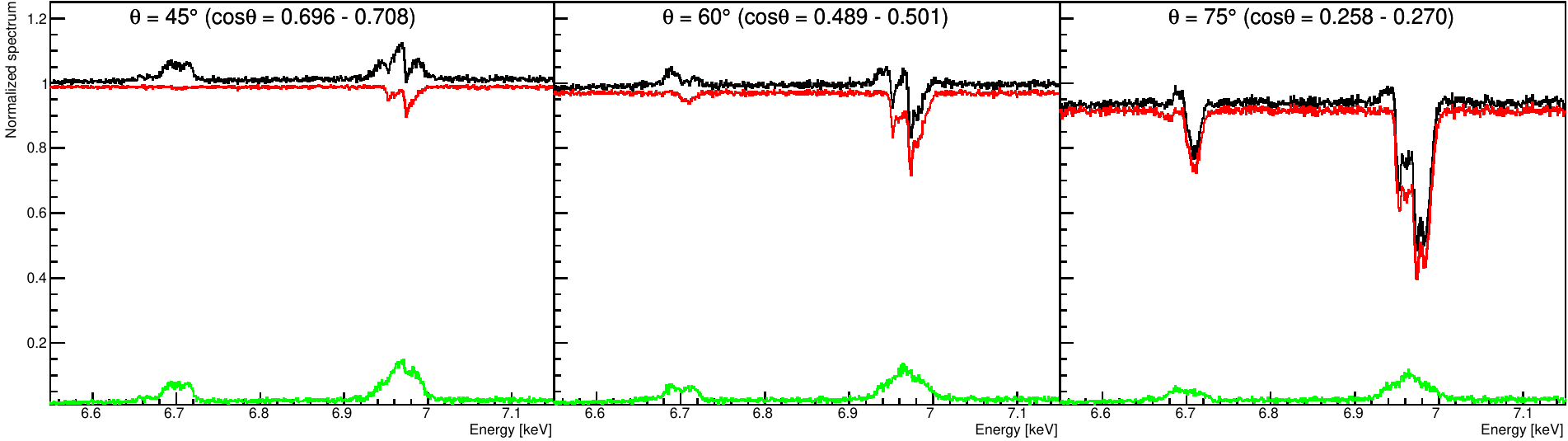}
    \caption{As in Fig.\ref{fig:vt_zero_each} but including additional saturated turbulent velocity 
    ($v_\text{turb} = v_{R}$) }
    \label{fig:vt_vr_each}
\end{figure*}


\begin{figure}
    \centering
    \includegraphics[width=0.9\hsize]{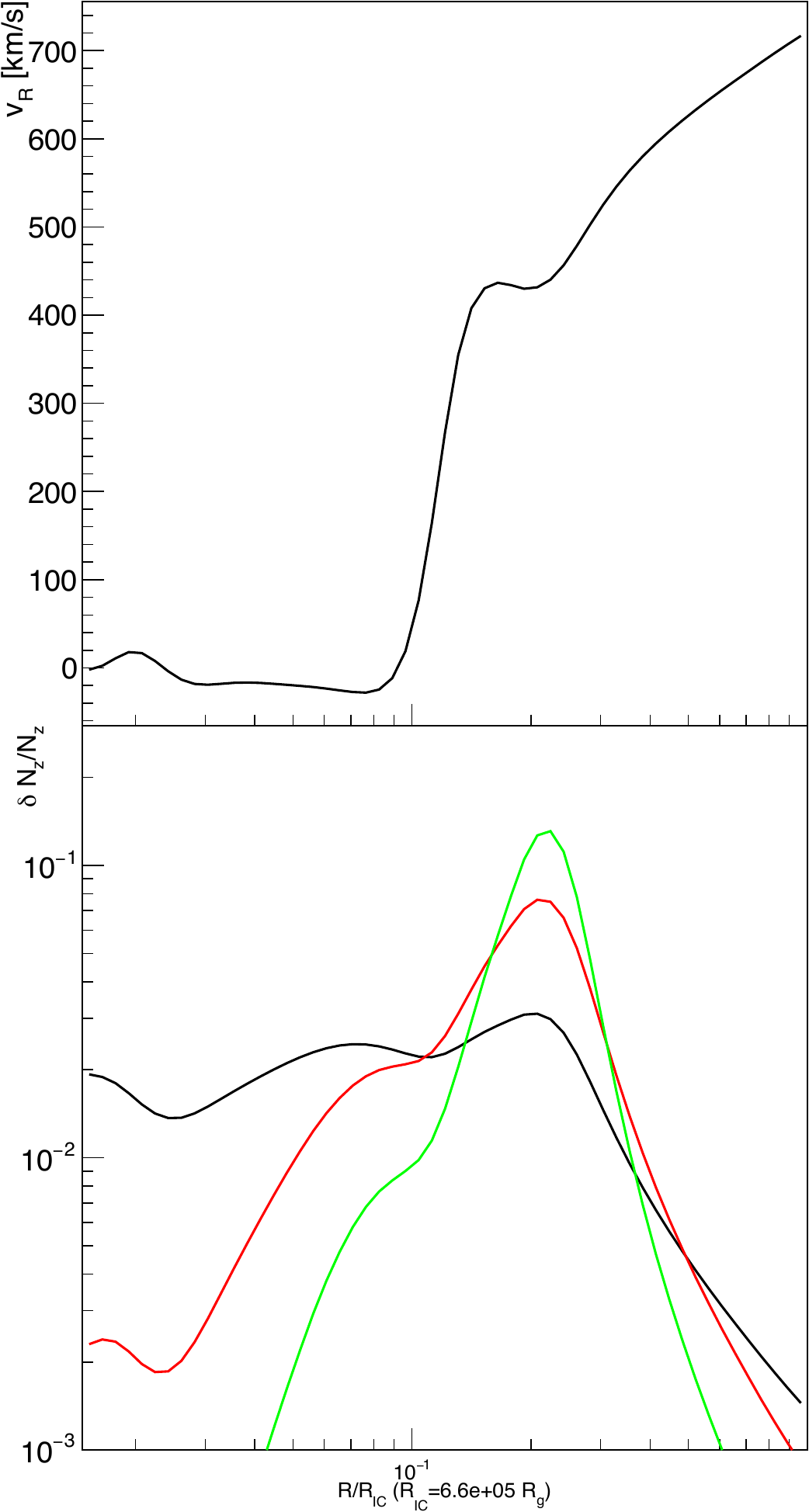}
    \caption{The radial profile of velocity (top) and ratios of local ion column density to that of total at $i = 75^\circ$ (bottom). 
    Colours show total column (black), Fe {\scriptsize XXVI} (red), and Fe {\scriptsize XXV} (green), respectively, }
    \label{fig:radial profile}
\end{figure}

\begin{figure}
    \centering
    \includegraphics[width=0.9\hsize]{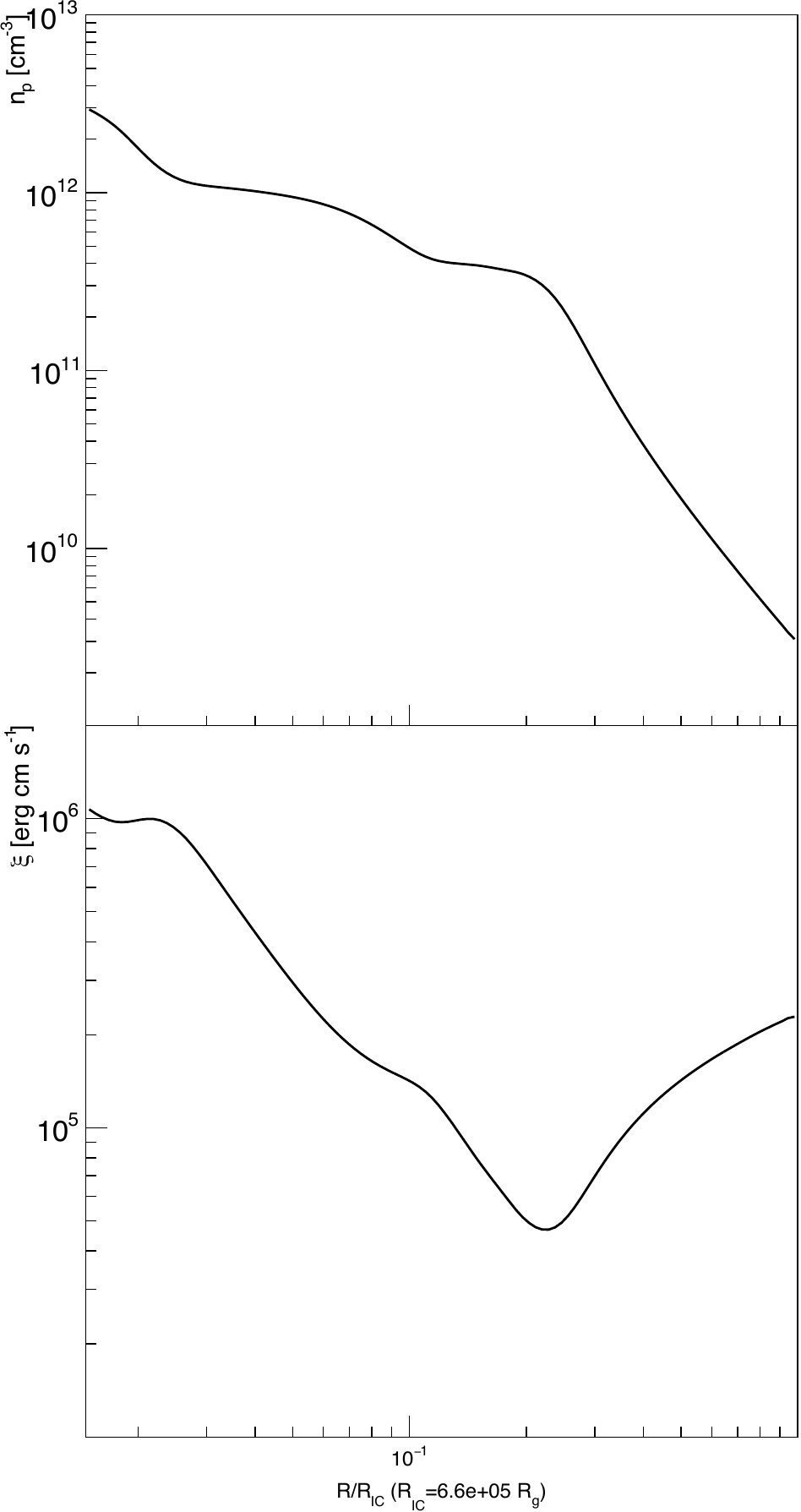}
    \caption{The radial profile of density (top) and ionisation parameter $\xi$ (bottom) at $i=75^\circ$.}
    \label{fig:radial_dens}
\end{figure}

We first simulate the spectrum exactly as predicted by the radiative hydrodynamic model. 
While this is the best calculation of the wind structure to date, there are still some 3D effects which it does not include, such as warping of the disc from radiation pressure and tidal forces \citep{Schandl1994,Ogilvie2001}, and the impact of the accretion stream onto the disc and wind \citep{Smak1970,Armitage1998}.
These processes could result in additional velocity components, which could develop into a fully turbulent flow, characterised by 
$v_\text{turb}=v_{R}$, so we show a second simulation including isotropic turbulence at this level. The only impact of this  turbulence is to give additional Doppler broadening to the emission and absorption lines.

We plot scattering/emission position distributions (Fig.\ref{fig:dist emission}). 
The scattering/emission occurs mostly at high inclinations,
reflecting the distributions of columns (total, H- and He-like iron). This shows
that our simulation box size is  large enough to capture all of the emission/scattering region.
These figures are made using the simulation without turbulence, but the total  scattering/emission 
probabilities are not dependent on turbulence.

Fig. \ref{fig:vt_zero_each} (without turbulence) and \ref{fig:vt_vr_each} (with turbulence) show the
resulting spectra at three different inclination angles. 
In both models, the absorption lines increase strongly with inclination angle as the line of sight intercepts more of the higher density material close to the disc plane, whereas the scattered 
component (sum of continuum plus the emission lines)
is almost constant with inclination angle.
Hence the emission lines are more prominent at low inclination angles as they are not filled in by the absorption. 

The model without turbulence shows that there are separate velocity components to the absorption lines at high inclinations. 
Fig.~\ref{fig:radial profile} shows the velocity structure along a line of sight at $75^\circ$, together with ion columns in  He- and H-like iron, and total.
The velocity plot clearly shows that there is static/inflowing material at small radii, at high ionisation state.
There is then a rapid acceleration zone from $R=0.1-0.2~R_\text{IC}$ as the thermal-radiative wind starts to be launched.
The wind is most efficiently launched at radii $\ge 0.2~R_\text{IC}$ so the wind mass loss rate rises towards larger radii, 
but its velocity also rises so the ion columns remain mostly constant in this region. The columns start to rise again as the velocity accelerates past 200~km/s and towards a plateau at 400~km/s, and the increase in density means a decrease in ionisation state, increasing the contribution from Fe {\scriptsize XXV}. Edge effects then come into play as 
the outer disc radius is reached at $R_\text{out}=0.2~R_\text{IC}$ for this simulation. The wind runs out of new material, so there is no back pressure from the wind outside of this point. The streamlines splay outwards, giving a fast drop in density, and hence a fast decrease in ion columns and increase in ionisation state (Fig.\ref{fig:radial profile}b). Thus the major part of both He- and H-like iron absorption arises from $\sim 0.2~R_\text{IC}$, where the material has a fairly constant velocity of around 400~km/s, with an  additional column of predominantly H-like Fe in the acceleration zone.
This explains the two sets of narrow Fe {\scriptsize XXVI} doublet lines seen in 
Fig.~\ref{fig:vt_zero_each}c.

To compare with previous analysis \citep{Miller2006b}, we plot the radial profile density and velocity distribution at the same inclination angle (Fig.\ref{fig:radial_dens}). 
Though the inner region has a large density, 
the ionisation parameter $\xi = L/(n_pR^2)~\mathrm{[erg~cm~s^{-1}]}$ of that material is large.  
As a result, columns of Fe {\sc xxvi} and {\sc xxv} are small in the inner region.
We mainly observe the region which has the largest ion columns at $\sim 0.2 R_\text{IC}$. 
In this region, the material has a density of $n_p \sim 3 \times 10^{11}~[\mathrm{cm^{-3}}]$, and an ionisation parameter of $\xi \sim 5\times 10^{4}$.
It is difficult to compare these with \citet{Miller2006b} as they do not give an equivalent ionisation parameter. Nonetheless, we are fitting the same data, and our models do fit. We can compare instead to \citet{Shidatsu2019}, who  used a one-zone model calculated by {\sc xstar} 
with the constant density of $n_p = 1.0\times 10^{12}$.
They obtained the ionisation parameter at the irradiation surface of absorber with $\xi = 3^{+2}_{-1} \times 10^{4}$,
which is consistent with our ionisation parameter at $R = 0.2 R_\text{IC}$.

Our assumption for the fully turbulent solution is that $v_\text{turb}=v_R$, so the turbulent velocity follows the radial outflow velocity structure.
This smears out some of the obvious velocity substructure, but the acceleration zone has lower turbulent broadening so is still distinct in the simulation. 

\begin{figure}
    \centering
    \includegraphics[width=0.9\hsize]{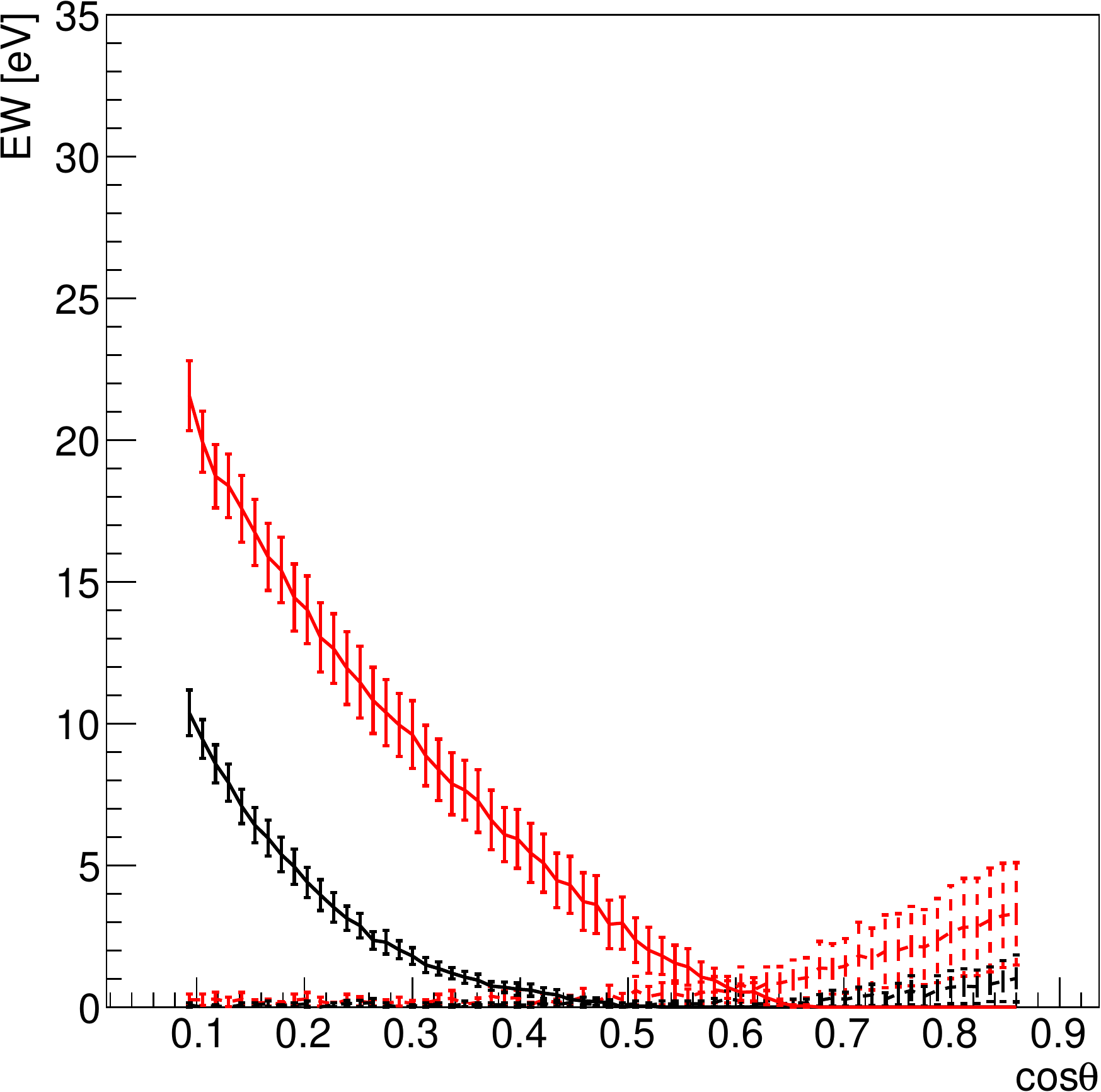}\\
    \includegraphics[width=0.9\hsize]{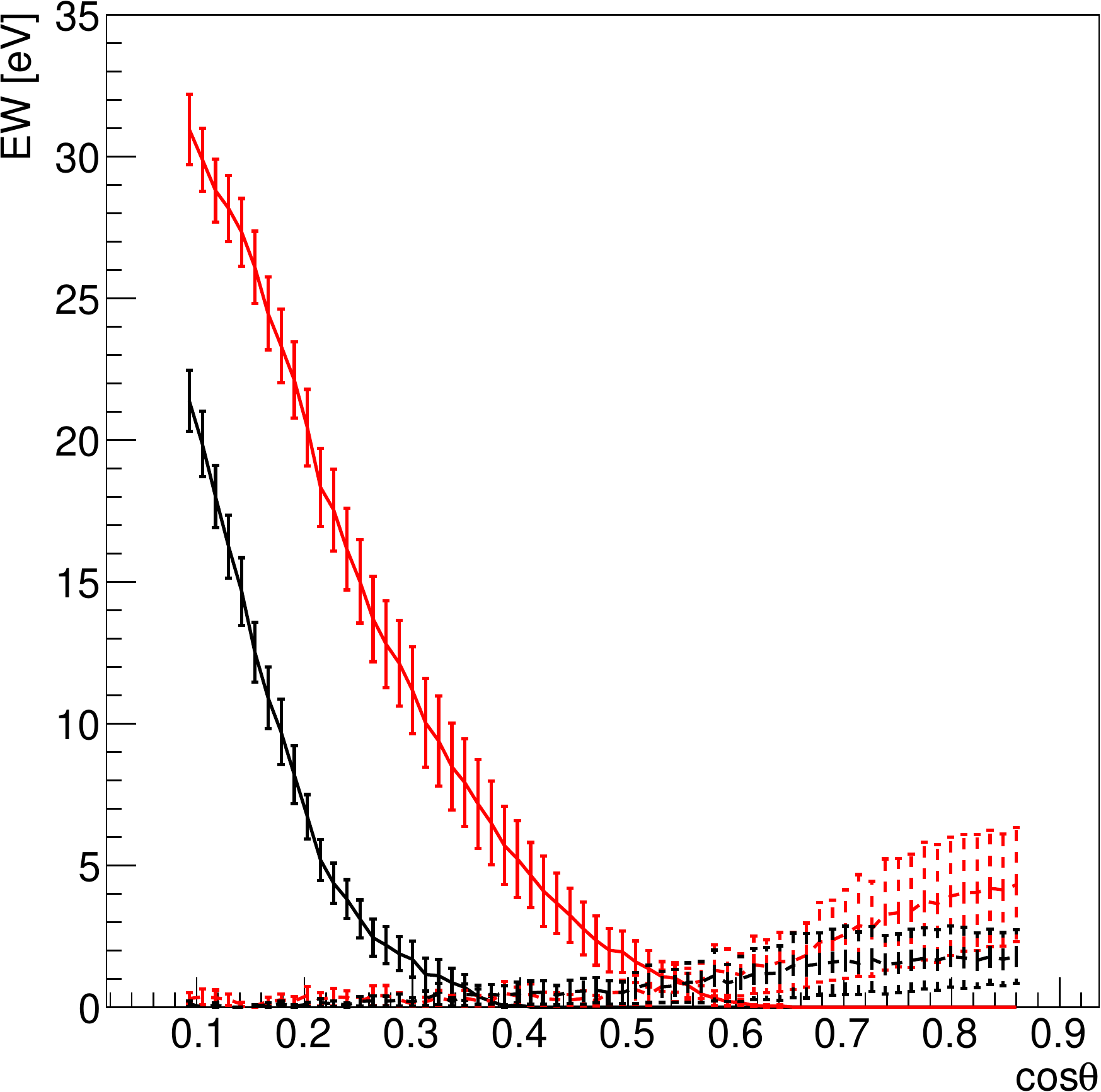}
    \caption{The angular dependence of absorption (solid) and emission (dashed) equivalent width Fe {\scriptsize XXVI} (Ly$\alpha_1$: 6.953 keV + Ly$\alpha_2$: 6.973 keV, red) and
    Fe {\scriptsize XXV} (He$\alpha~y$: 6.668 keV + He$\alpha~w$: 6.700 keV, black ) with $v_\text{turb} = 0$ (top) and $v_\text{turb} = v_{R}$ (bottom).
    Error bars are calculated from Poisson noise.}
    \label{fig:EW}
\end{figure}

We show the equivalent width (EW) of the H- and He-like
absorption and emission lines at each angle in Fig.~\ref{fig:EW}.
We measure these following \citet{Tomaru2018} by
fitting the total spectrum (black lines in Fig.~\ref{fig:vt_vr_each}) 
by an arbitrary function $aE^2+bE+c$ ($a, b$ and $c$ are free parameters), excluding the line regions,
then numerically integrate the difference between this continuum and the simulation data. 
EWs for the model including turbulence are larger at the highest inclination angles as the lines in the initial simulation are saturated. 
The maximum Fe {\scriptsize XXVI} EW is around 
$\sim 30~\text{eV}$ (with turbulence) compared to 
$\sim 20~\text{eV}$ (without). 
The EWs of emission are small,
approaching 5 eV only at face on inclination angle.

\section{Fitting to current and future high resolution data}

\subsection{{\it Chandra}/HETGS}

\begin{figure*}
    \centering
    \includegraphics[width=0.45\hsize]{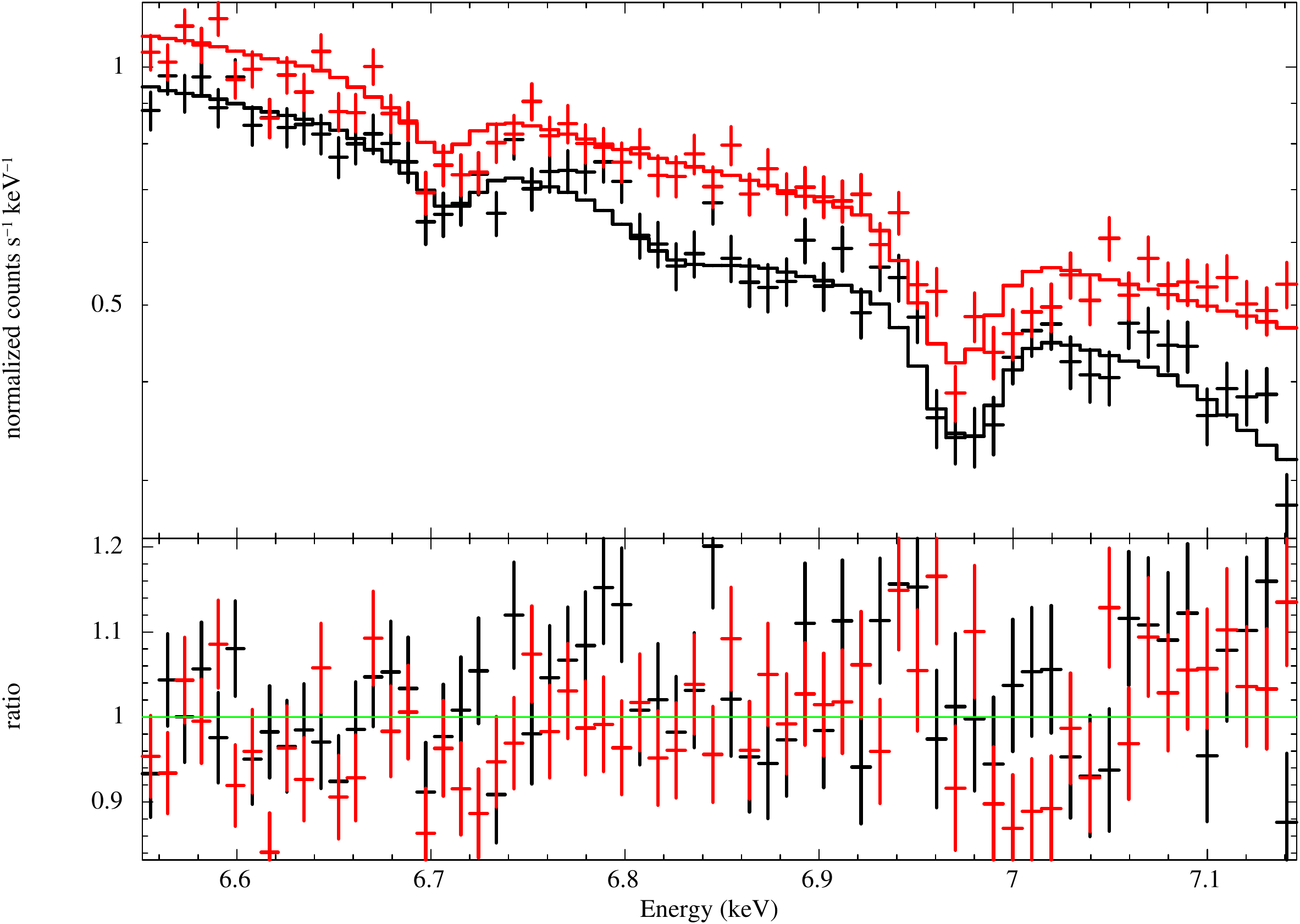}
    \includegraphics[width=0.45\hsize]{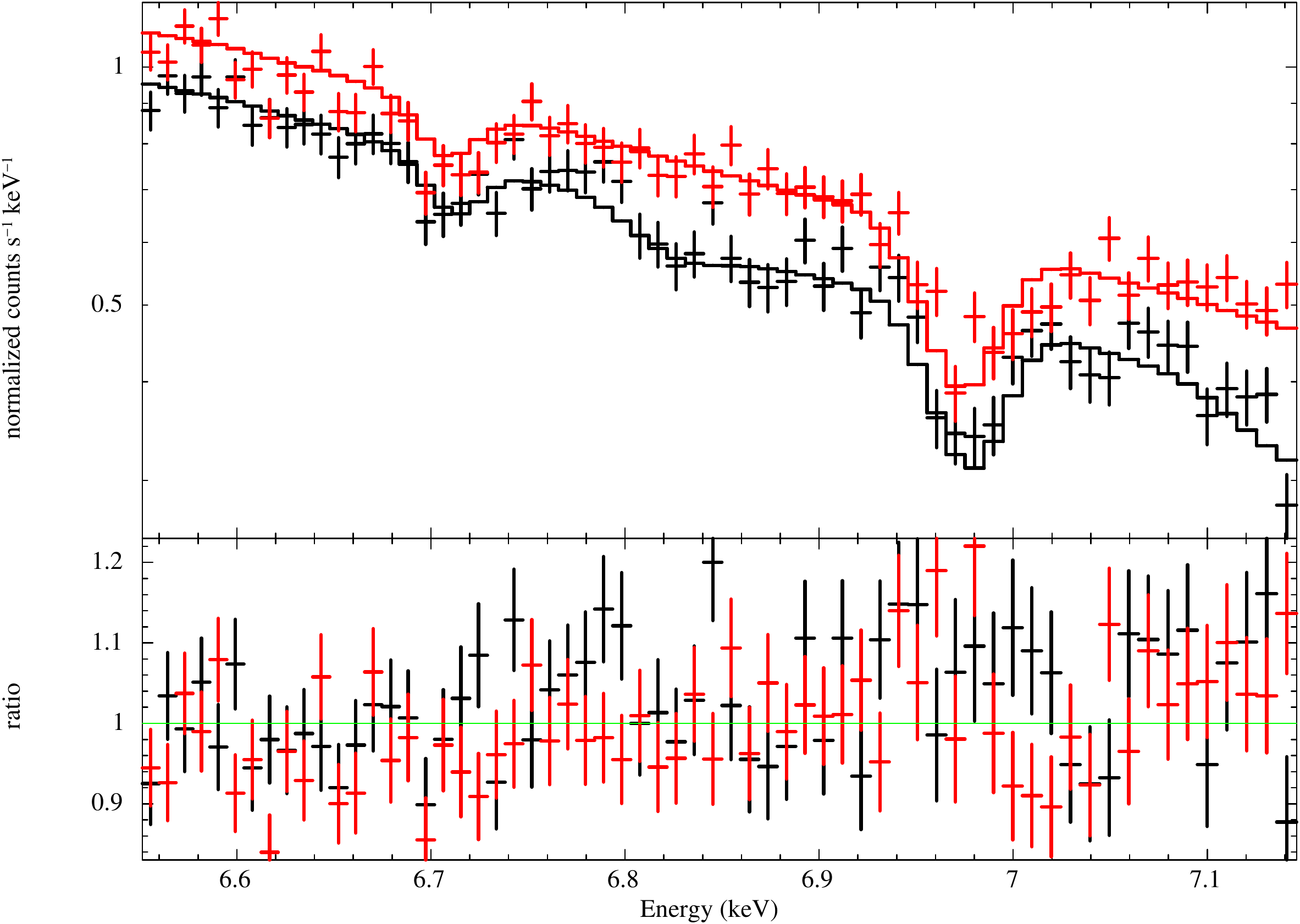}
    \caption{{\it Chandra}/HETGS data with best fit model of $v_\text{turb} = 0$ (left) and $v_\text{turb} = v_R$ (right). Best fit inclination angle is $81^{\circ}$ and $78^{\circ}$ respectively.}
    \label{fig:model_fit}
\end{figure*}

We now take the {\sc monaco} results both without and with turbulence and fit them to the {\it Chandra} high energy transmission grating spectrometer (HETGS) first order data of the soft state in H1743-322 (OBSID:3803).
Free parameters in these fits are inclination angle and normalization (We implement these models into {\sc XSPEC} as tabulated additive model which include the continuum component because emission depends on continuum.).
Both models can fit the  observational data equally well (Fig.~\ref{fig:model_fit}) for inclination angles of $81^\circ$ (without turbulence) or $78^\circ$ (with turbulence).
This could indicate a marginal preference for turbulence as the expected angle from the binary parameters is $75\pm 3^\circ$ \citep{Steiner2012}, but the difference is small and the difference in $\chi^2$ has more to do with the line equivalent width (which could be changed easily by changing the disc size and/or source luminosity: Paper I) than with the intrinsic width. 

To really resolve the line profiles such as the velocity separation of static corona and the outflowing wind and understand whether or not turbulence is present requires higher resolution than possible in {\it Chandra} grating data. Hence we now 
consider whether the X-ray micro calorimeters on-board future X-ray satellites such as {\it XRISM} can better distinguish the intrinsic velocity structure.

\subsection{Simulated {\it XRISM}/Resolve observations} 

We simulate a {\it XRISM}/Resolve observation from the 
simulation models which give the best fit to the {\it Chandra}/HETGS data. The Resolve calorimeter gives 5~eV energy resolution at 6~keV in its best modes (High plus Medium resolution events, hereafter H+M). However, this energy resolution is produced by modelling the pulse height shape to determine the total energy of the event. This is done by template matching to the fast rise, exponential decay shape, and the energy in the event can be accurately reconstructed if no other event arrives in the 30~millisecond period covered by the template. Where events overlap, the energy cannot be reconstructed so accurately (Low resolution events). 
With higher sampling rate it might be possible 
to determine the pulse shape more accurately for these piled-up events, but the on-board computer power is limited and can only handle template matching for a maximum 50~c/s. Above this limit, Low resolution events give a degraded resolution of around 30~eV at iron (comparable to {\it Chandra}/HETGS first order data). 

The {\it XRISM}/Resolve detector is split into four quadrants, with counts from each quadrant handled separately, so the maximum count rate is 200~c/s, which corresponds to 
around 100~mCrab source for an on-axis point source. 
Sources with higher predicted count rates will need mitigation strategies for the observations to reduce the count rates to these levels, using a combination of filters and/or offsets.
These effects can be seen in the 
{\it Hitomi}/SXS observation of the Crab. This was performed with the gate valve closed, which cuts out all flux below 2~keV and reduces the higher energy count rate by a factor 2 compared to a standard observation. These SXS data did indeed show a $\sim 50$~c/s/quadrant limit ({\it Hitomi} collaboration 2017)

We simulated the soft spectrum of H1743-322 from a continuum model determined from the quasi-simultaneous RXTE data corresponding to the {\it Chandra}/HETGS data (OBSID:3803).
This has 2-10~keV of $1.7\times 10^{-8}$~ergs~cm $^{-2}$ ~s $^{-1}$, or around 1 Crab, and predicted 
0.3-2~keV flux of $2.3\times 10^{-9}$~ergs~cm$^{-2}$~s$^{-1}$ using an assumed column density of $1.6\times 10^{22}$~cm$^{-2}$ \citep{Shidatsu2019}.
This results in 1000~c/s, clearly outside of the scope of the electronics, so we consider mitigation strategies. 

The Beryllium filter is designed especially to reduce the low energy 
flux so as to leave as many counts as possible in the higher energy 
bandpass. However, interstellar absorption in our Galaxy produces a 
similar effect. It makes only a 20\% reduction in the count rate of 
H1743-322, though all the counts lost are soft. The effect is similar to a neutral column density of $2.5\times 10^{22}$~cm$^{-2}$, so this will not have much effect for more absorbed bright Galactic sources. 
The ratio of flux in the iron line band (6.5-7.5~keV) to the full band (0.3-10~keV) is $\sim 1/40$ without the Be filter, and $\sim 1/33$ with the Be filter. A combination of the Be filter and an 
offset pointing, so that the image is centred on one quadrant rather than illuminating all quadrants equally, may completely saturate one quadrant, but allow events from the other 3 quadrants to be close to the maximum, giving H+M events. The total H+M high resolution events could then be up to 150~c/s, though 100~c/s is a more reasonable 
expectation factoring in potential losses from cross-talk. 

\begin{figure*}
    \centering
    \includegraphics[width=0.45\hsize]{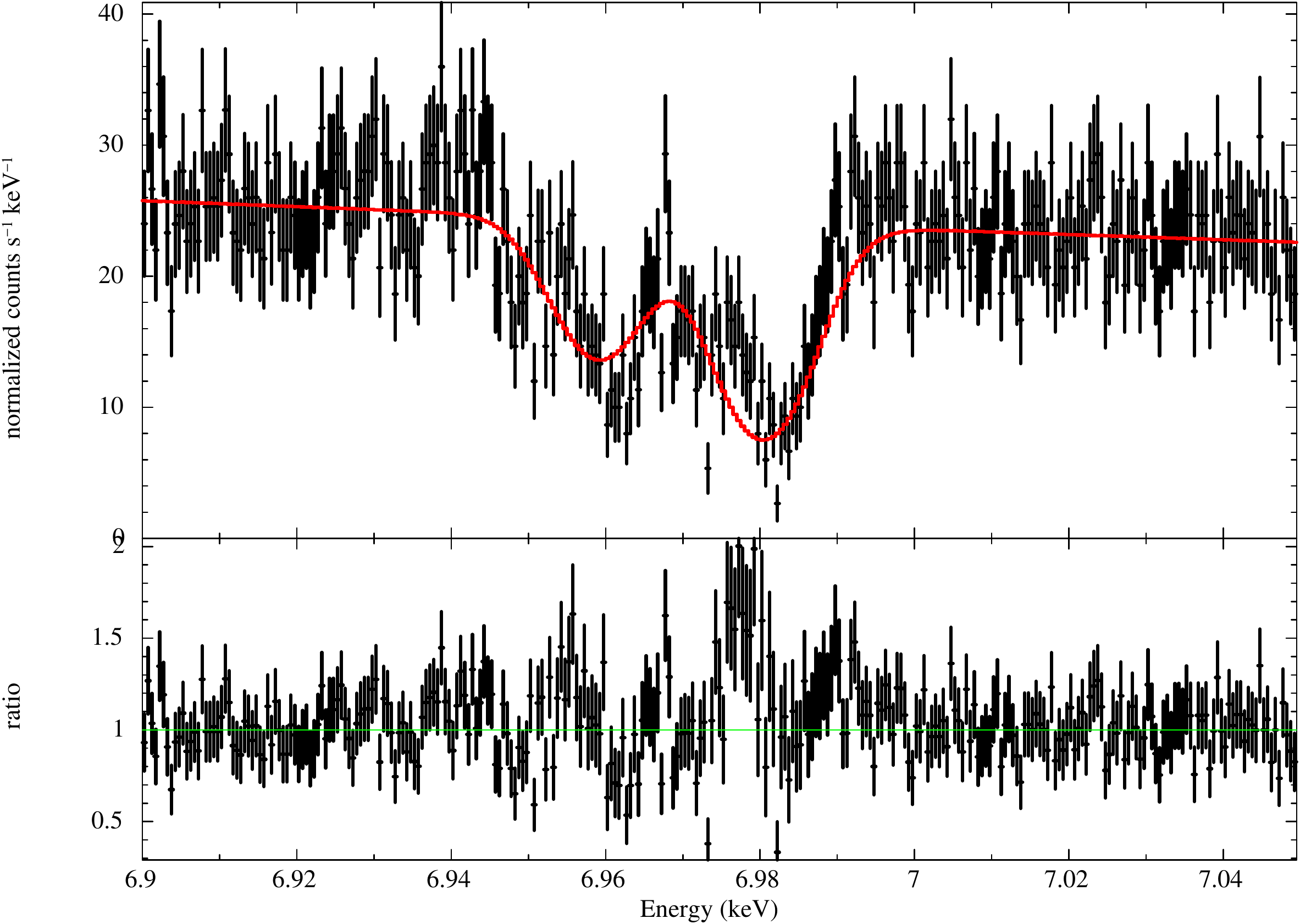}
    \includegraphics[width=0.45\hsize]{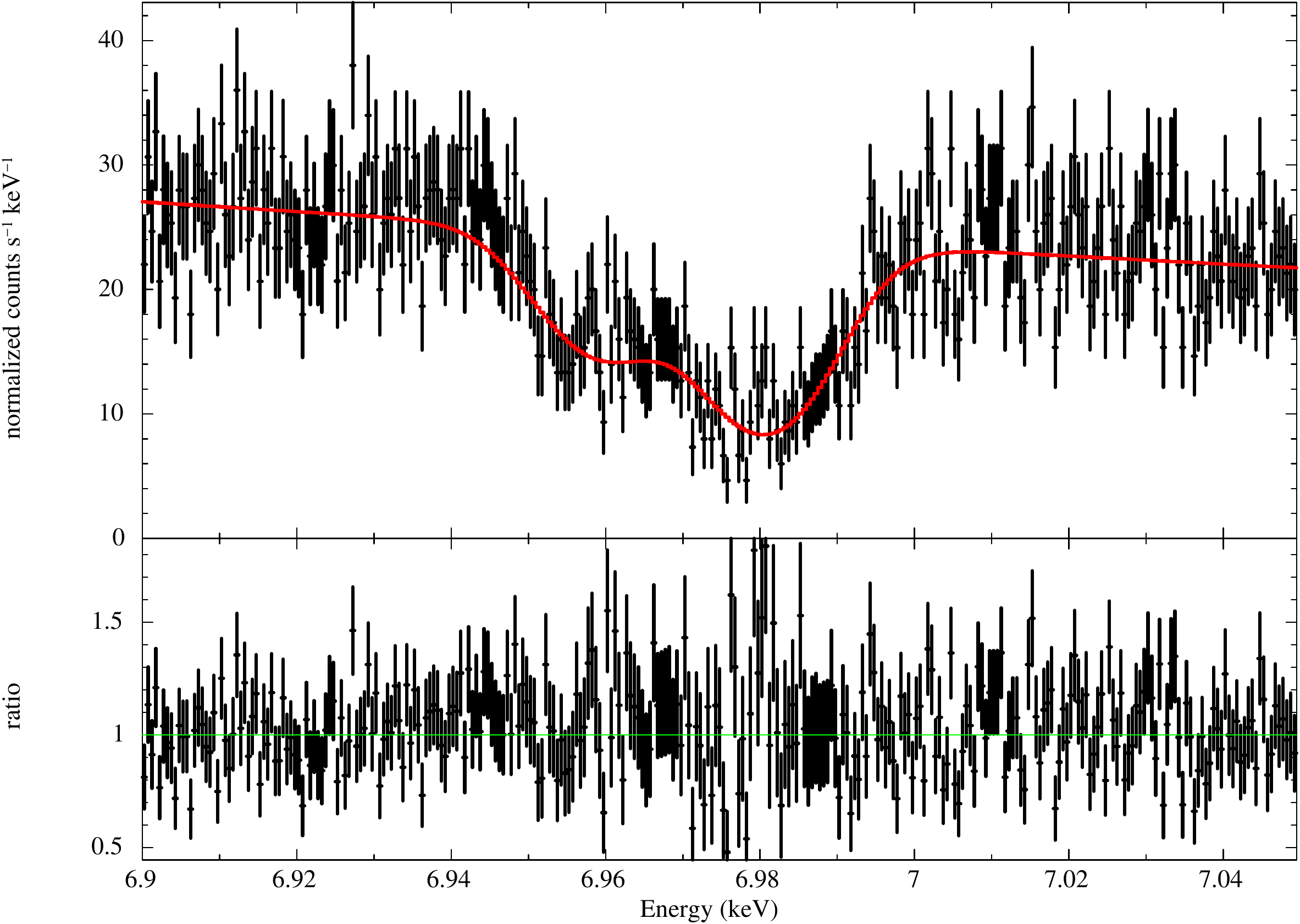}
    \caption{Simulated spectrum of a 30ks {\it XRISM} observation (live time fraction of 0.1) for the model with no additional turbulence (left), and $v_\text{turb} = v_R$ (right) around the Fe {\scriptsize XXVI} K$\alpha$ doublet line energy.
    The absorption is modelled using a single {\sc kabs} component (red). The lower panel shows the residuals as a ratio between the data and model at each energy. Clearly Resolve can determine the velocity structure in the lines even at these low velocities.}
    \label{fig:3ks obs Kabs}
\end{figure*}

\begin{figure*}
    \centering
    \includegraphics[width=0.45\hsize]{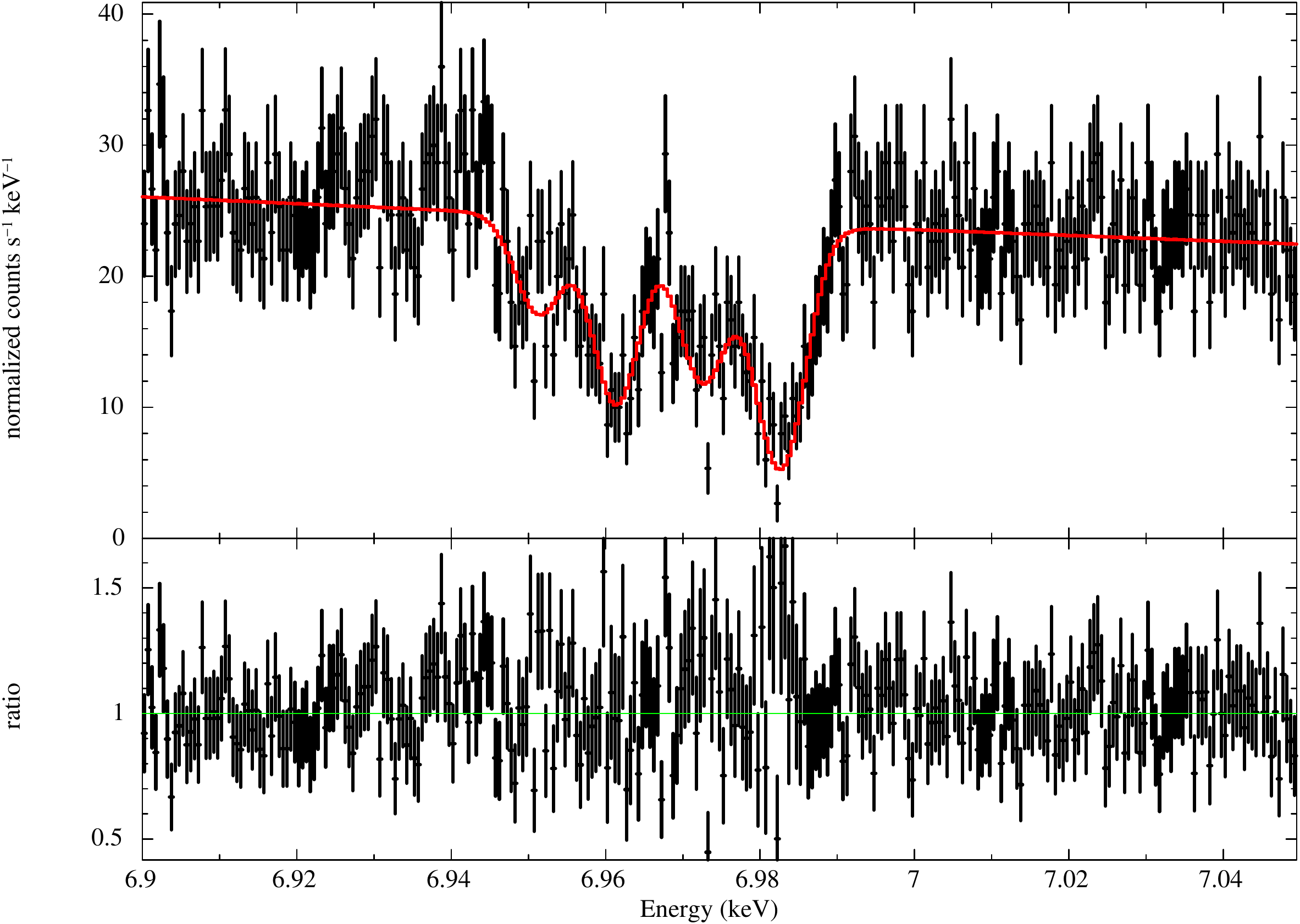}
    \includegraphics[width=0.45\hsize]{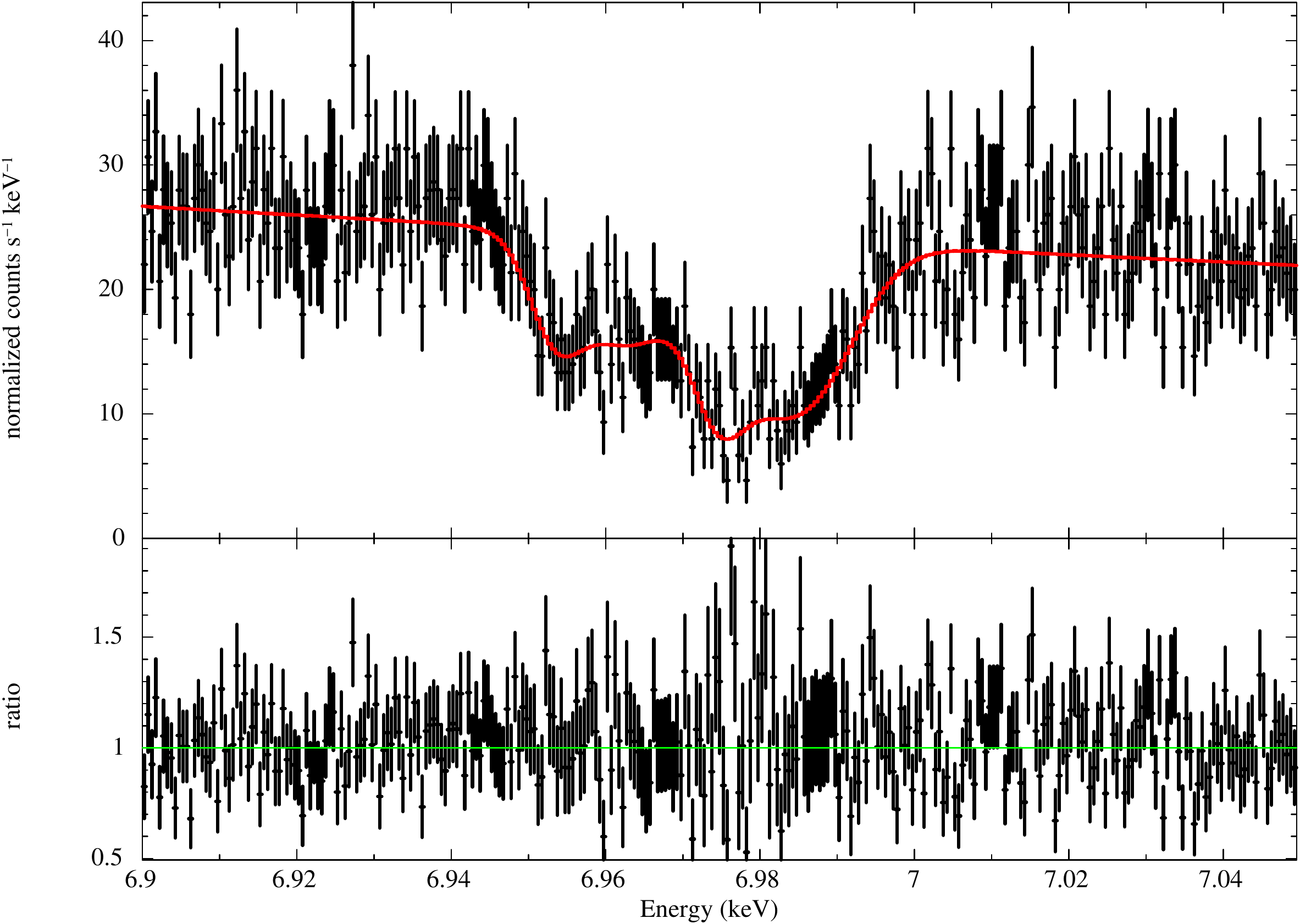}
    \caption{As Fig.~\ref{fig:3ks obs Kabs} but fitted by double {\sc kabs} model}
    \label{fig:2Kabs}
\end{figure*}

Alternatively, there is a neutral density filter which reduces flux at all energies by a factor 4. This would reduce our predicted count-rate to $\sim 250$~c/s not far from the electronics limit of 
$\sim 200$~c/s. 

Hence we expect a maximum count rate of 100-200~c/s for the H+M resolution data from our initial spectrum which has 1000~c/s.
We conservatively assume a factor 10 loss,
so we simulate a 3~ks {\it XRISM}/Resolve exposure as corresponding to a 30~ks observation. 
Fig.~\ref{fig:3ks obs Kabs} shows the resulting energy spectrum and its statistical uncertainties around the Fe {\scriptsize XXVI} iron line band for the model without turbulence and then one with turbulence. 
We fit both spectra with a single  {\sc kabs} component (\citealt{Ueda2004}, including corrections from \citealt{Kubota2007}). This model
includes the full Voigt line profile for both components of the doublet (K$\alpha_1$ and K$\alpha_2$, with ratio fixed to that expected from atomic physics). The free parameters are
the column density of the ion, the isotropic turbulent velocity, in terms of the equivalent temperature, $kT$~keV, and any red or blue shift indicating bulk inflow/outflow. 
The lower panel shows the residuals to this fit, with parameters detailed in Table~\ref{tab:single}.
    
\begin{table}
\centering
\caption{Single {\sc kabs} fits to the simulated {\it XRISM}/Resolve data}
\label{tab:single}
\begin{tabular}{llc}
\hline
\hline
& $v_\text{turb}=0$ & $v_\text{turb}=v_{R}$ \\
\hline
$N_\text{XXVI}~ (10^{18}~\text{cm}^{-2})  $& $0.55\pm 0.04$ &  $0.67\pm 0.05$ \\
$kT$ (keV)          & $30_{-5}^{+6}$      & $59_{-10}^{11}$ \\
$z\times 10^{-3}$ &$-1.04\pm 0.08$ & $-1.1\pm 0.1$ \\
$\chi^2/\nu$      & 350/294 & 326/294\\
\hline
\end{tabular}
\end{table}
    
\begin{table}
\centering
\caption{Double {\sc kabs} fits to the simulated {\it XRISM}/Resolve data}
\label{tab:double}
\begin{tabular}{llc}
\hline
\hline
& $v_\text{turb}=0$ & $v_\text{turb}=v_R$ \\
\hline
$N_\text{1,XXVI}~(10^{18}~\text{cm}^{-2})$   & $0.46_{-0.06}^{+0.14}$   & $0.49_{0.18}^{+0.11}$\\
$kT_1$ (keV)        & $3.4_{-1.8}^{+2.3}$      & $43 \pm 22 $  \\ 
$z_1\times 10^{-3}$ & $-1.35\pm 0.05 $         & $-1.5_{-0.4}^{+0.3}$ \\
$N_\text{2, XXVI}~(10^{18}~\text{cm}^{-2})$  & $0.18\pm 0.03$           & $0.17_{-0.09}^{0.17}$ \\ 
$kT_2$ (keV)        & $4.4_{-3.1}^{+5.7}$      & $5.4\pm^{+11}_{-5.0} $     \\ 
$z_2\times 10^{-3}$ &   $0.08_{-0.11}^{+0.09}$ & $-0.2_{-0.2}^{+0.1}$ \\
$\chi^2/\nu$        & 261/291                  & 308/291 \\
\hline
\end{tabular}
\end{table}

There are clear residuals in the pure hydrodynamic simulations results. We include a second {\sc kabs} component and the fit is very significantly improved, with $\Delta\chi^2=90$ for 3 additional free parameters. There is a smaller but still significant improvement for 
the turbulent wind, with $\Delta\chi^2=20$, with all parameters of the two components shown in Table~\ref{tab:double}. This shows that
{\it XRISM} can resolve the infall/outflow velocity structure expected even in the fairly weak thermal-radiative wind modelled here, and even in the presence of fully developed turbulence. 

\begin{figure}
    \centering
    \includegraphics[width=0.9\hsize]{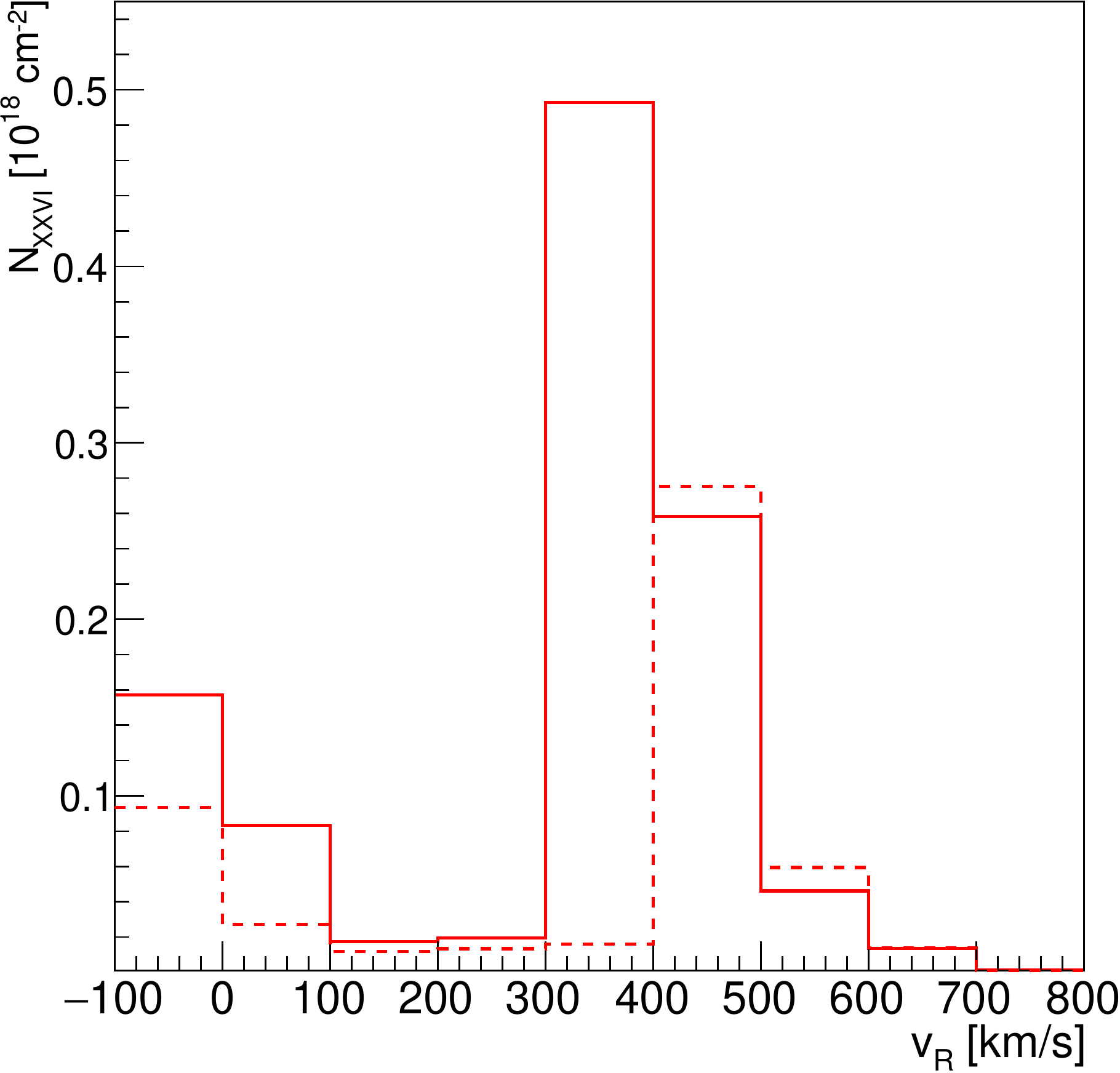}
    \caption{The column density distribution of Fe {\sc xxvi} as a function of outflow velocity at $75^\circ$ (dashed) and $80^\circ$ (solid).
    }
    \label{fig:vr_wc}
\end{figure}

To confirm whether double {\sc kabs} can reproduce the column density of hydrodynamic result, 
we plot the column density distribution of Fe {\sc xxvi} as a function of outflow velocity (Fig. \ref{fig:vr_wc}) at $80^\circ$ (solid) and $75^\circ$ (dashed).
The obtained Fe {\sc xxvi} columns in Tab.\ref{tab:double} well match peak values of this column distribution around 0 km/s and 400 km/s at $80^\circ$.

\section{Discussion: distinguishing between thermal-radiative and magnetic winds}

The {\it XRISM} simulations shown here can clearly resolve the line produced in this thermal-radiative simulation into two velocity substructures, one which is static (or slightly infalling) from the X-ray heated disc atmosphere at small radii where the material does not have enough thermal energy to escape, and the outflowing wind at larger radii.
This can be seen even in the presence of saturated turbulence.
Conversely, current models of magnetically driven winds (e.g.\citealt{Fukumura2010, Chakravorty2016}) which assume self-similar magnetic fields from the entire disc launch winds at all radii, with velocity decreasing going outwards. 
These models do not predict that there can be static material anywhere, least of all at radii smaller than where the outflow is produced. 
However, this static material in the thermal-radiative winds can only be seen if there is substantial column in Fe {\scriptsize XXVI} and {\scriptsize XXV} at the launch radius.
A much higher luminosity could result in material close to the launch radius being completely ionised, so that Fe {\scriptsize XXVI} and {\scriptsize XXV} are predominantly produced at larger radii so would not show this characteristic signature.
Alternatively, a much larger disc but with the same irradiating spectrum and luminosity would produce more wind, so the contribution of the static material might be lost in the heavier absorption from the outflow.
H1743-322 in this soft state ($L \sim 0.3L_\text{Edd}$ and $T_\text{IC,8}\sim 0.10$) has the ideal parameters to show this transition from static corona to wind predicted in the thermal-radiative models

However, we note that testing this characteristic velocity pattern is already possible using the neutron star binary systems.
The sample accumulated by \citet{DiazTrigo2016} shows that winds are only seen in systems with large discs, whereas systems with small discs have only static absorption features.
Thus it is already clear that the absorbing material forms a structure where there is static material at smaller radii, with the wind produced only at larger radii.
This is clearly consistent with expected behaviour of thermal-radiative winds as stressed by \citet{DiazTrigo2016}.
However, here we stress the converse, that this is {\em inconsistent} with self-similar magnetic wind models, 
since they are outflowing everywhere, and with larger velocity at smaller radii. 
These models predict that systems with smaller discs which show absorption features should have faster winds. 
This is not observed, ruling out a origin of self-similar magnetic winds for these features. 

 There could still be magnetic winds which are important dynamically, but are too highly ionised to give absorption features.
 Large scale magnetic fields torque the disc, so can be the source of angular momentum transport even if the resulting Blandford-Payne wind \citep{Blandford1982} is not the source of the observed absorption features. 
 However, there are also problems with this approach as a self-similar magnetic wind is (by definition) launched at all radii, so it will disrupt the static disc atmosphere at small radii predicted by the thermal wind models, and seen in the small disc systems.
 This atmosphere (and wind in the larger system) acts as a calorimeter, showing how much energy and momentum is transported by an otherwise invisible magnetic wind.
 While the thermal-radiative wind region has dense material preferentially in equatorial directions, the X-ray heated disc atmosphere lies directly above the disc so is sensitive to energy/momentum flux from a magnetic wind launched in any direction.
 The extent to which {\it XRISM}/Resolve observations show consistency with the thermal-radiative disc atmosphere/wind predictions also strongly constrains the existence of self-similar magnetic wind from the disc. 

One of magnetic winds not constrained by these considerations is one which arises only in the hot inner flow region.
This is characteristic of the truncated disc models for the low/hard state in black hole binaries (equivalently the island state in neutron stars).
This region is radially separated from the disc, so a Blandford-Payne wind anchored into this inner hot flow would not impact and/or disrupt the X-ray heated atmosphere/wind over the disc.
However, a more common name for such a wind is the jet, and the best current models which use large scale magnetic fields to transport angular momentum in the hot flow region are focused on powering the compact radio jets seen in this state \citep{Marcel2018a, Marcel2018b}. 

\section{Conclusions}

In our previous work we showed state of the art numerical hydrodynamic simulations of a thermal-radiative wind.
We tailored this simulation to explain the column density, ionisation state and velocity of the absorption features seen in {\it Chandra}/HETGS data from the black hole binary system H1743-322 in its soft, disc dominated state.
Here we calculate the detailed absorption and emission line profiles predicted by this model by using the density/velocity structure as input to a Monte Carlo radiation transport code.
The results fit well to the HETGS data, showing that these physical wind models can indeed be the origin of the absorption features seen, rather than requiring a magnetically driven wind. 

This is the best current calculation of these winds and their observable absorption/emission features, but it does not incorporate all the potential physics of the outer disc.
This could include warps and the impact of the accretion stream, both of which could induce turbulence, so we also calculate the line profile predicted for saturated turbulence in the thermal-radiative wind.
Current HETGS data can neither distinguish models with and without additional turbulence nor identify the velocity separation between the static corona and the outflowing wind, but we show that future observations with {\it XRISM}/Resolve (due for launch Jan 2022) will be able to distinguish the detailed velocity structure in the absorption lines at this level, giving an unprecedented view of the wind launch and acceleration processes.

While it is already clear that
thermal-radiative winds are consistent with
current observations of H1743-322, 
using the entire sample of absorption features detected from accreting black holes and neutron stars gives more information. 
These show that static absorbers are seen in systems with small discs, whereas winds are only seen in larger systems. 
This is not only  consistent with thermal-radiative winds, but crucially, is inconsistent with self-similar magnetic winds as these instead predict faster outflows at smaller radii.
Thus, these magnetic winds are strongly disfavoured as the origin of the absorption features seen in the black hole and neutron star binary systems.




\section*{Acknowledgements}

We acknowledge the work of the Hitomi SXS and XRISM Resolve teams in understanding bright source observations, especially M. Tsujimoto, J. Miller and E. Hodges-Kluck.
This work supported by JSPS KAKENHI Grant Number JP 19J13373 (RT), 
Society for the Promotion of Science Grant-in-Aid for Scientific Research (A) (17H01102 KO; 16H02170 TT), 
Scientific Research (C) (16K05309 KO; 18K03710 KO), 
and Scientific Research on Innovative Areas (18H04592 KO; 18H05463 TT).
This research is also supported by the Ministry of Education, Culture, Sports, Science and Technology of Japan as "Priority Issue on Post-K computer"(Elucidation of the Fundamental Laws and Evolution of the Universe) and JICFuS. 
RT acknowledges the support by JSPS Overseas Challenge Program for Young Resarchers.
CD acknowledges the Science and Technology Facilities Council (STFC) through grant ST/P000541/1, and visitor support from Kavli IPMU  supported in part by the National Science Foundation under Grant No. NSF PHY-1748958.




\bibliographystyle{mnras}
\bibliography{library} 





\bsp	
\label{lastpage}
\end{document}